\documentclass[%
 reprint,
superscriptaddress,
 amsmath,amssymb,
aps,
]{revtex4-1}
\usepackage[utf8]{inputenc}
\usepackage{xcolor}
\usepackage{amsmath}
\usepackage{graphicx}
\usepackage{booktabs}
\usepackage{hyperref}
\hypersetup{
    colorlinks=true,
    linkcolor=blue,
    filecolor=magenta,      
    urlcolor=red,
    citecolor=blue,
}

\usepackage{graphicx}% Include figure files
\usepackage{dcolumn}% Align table columns on decimal point
\usepackage{bm}% bold math

\begin{document}

\preprint{APS/123-QED}

\title{Probing d-wave superconducting gap of high-$T_\mathrm{c}$ cuprate $\mathrm{Bi}_2\mathrm{Sr}_2\mathrm{Ca}_2\mathrm{Cu}_3\mathrm{O}_{10+\delta}$ by resonant inelastic X-ray scattering}

\author{Kunhao Li}
\affiliation{International Center for Quantum Materials, School of Physics, Peking University, Beijing 100871, China}

\author{Qizhi Li}
\affiliation{International Center for Quantum Materials, School of Physics, Peking University, Beijing 100871, China}

\author{Changwei Zou}
\affiliation{International Center for Quantum Materials, School of Physics, Peking University, Beijing 100871, China}

\author{Jaewon Choi}
\affiliation{Diamond Light Source, Harwell Campus, Didcot, UK}

\author{Chaohui Yin}
\affiliation{Beijing National Laboratory for Condensed Matter Physics, Institute of Physics, Chinese Academy of Sciences, Beijing 100190, China}

\author{Mirian Garcia-Fernandez}
\affiliation{Diamond Light Source, Harwell Campus, Didcot, UK}

\author{Stefano Agrestini}
\affiliation{Diamond Light Source, Harwell Campus, Didcot, UK}

\author{Shilong Zhang}
\affiliation{International Center for Quantum Materials, School of Physics, Peking University, Beijing 100871, China}

\author{Chengtian Lin}
\affiliation{Max Planck Institute for Solid State Research, Stuttgart, Germany}

\author{Xingjiang Zhou}
\affiliation{Beijing National Laboratory for Condensed Matter Physics, Institute of Physics, Chinese Academy of Sciences, Beijing 100190, China}

\author{Ke-Jin Zhou}
\affiliation{Diamond Light Source, Harwell Campus, Didcot, UK}
\affiliation{School of Nuclear Science and Technology, University of Science and Technology of China, Hefei 230026, China}

\author{Yi Lu}
\affiliation{National Laboratory of Solid State Microstructures and Department of Physics, Nanjing University, Nanjing 210093, China}
\affiliation{Collaborative Innovation Center of Advanced Microstructures, Nanjing University, Nanjing 210093, China}

\author{Yingying Peng}
\email{yingying.peng@pku.edu.cn}
\affiliation{International Center for Quantum Materials, School of Physics, Peking University, Beijing 100871, China}
\affiliation{Collaborative Innovation Center of Quantum Matter, Beijing 100871, China}

\date{\today}

\begin{abstract}

The superconducting gap is a characteristic feature of high-T$_c$ superconductors and provides crucial information on the pairing mechanism underlying high-temperature superconductivity. Here, we employ high-resolution resonant inelastic X-ray scattering (RIXS) at the Cu $L_3$-edge to investigate the superconducting gap in the overdoped cuprate $\mathrm{Bi}_2\mathrm{Sr}_2\mathrm{Ca}_2\mathrm{Cu}_3\mathrm{O}_{10+\delta}$ ($T_\mathrm{c}$ = 107 K). By analyzing antisymmetrized, temperature-dependent RIXS spectra over a range of in-plane momentum transfers, we observe a clear suppression of low-energy spectral weight below T$_c$, indicative of superconducting gap formation. This suppression is most pronounced at small momentum transfers ($|\boldsymbol{q}_\parallel| \leq 0.18$ r.l.u.) and corresponds to a gap size of approximately 2$\Delta_0 \sim$ 130 meV. Comparison with theoretical calculations of the momentum-dependent charge susceptibility supports a d-wave symmetry of the superconducting gap, while an isotropic s-wave gap fails to reproduce key experimental features. These findings establish RIXS as a powerful, bulk-sensitive probe of superconducting gap symmetry and highlight its utility for studying materials beyond the reach of surface-sensitive techniques such as ARPES and STM.

\end{abstract}

\maketitle

\section{\label{sec:intro}Introduction}

Cuprate superconductors exhibit the highest critical superconducting temperature ($T_\mathrm{c}$) at ambient pressure, which is significantly higher than those of conventional superconductors \cite{Bednorz1986,YBCO1987,HBCO1993}, establishing them as a subject of enduring interest in both fundamental research and technological applications. Cuprates possess quasi-two-dimensional layered crystal structures, strong electronic correlations, and complex phase diagrams \cite{cupratesgap}. Among the cuprate family, the bismuth-based cuprates have been extensively investigated\,\cite{BSCCO} because it can be easily cleaved for surface-sensitive measurements. 
The trilayer  $\mathrm{Bi}_2\mathrm{Sr}_2\mathrm{Ca}_2\mathrm{Cu}_3\mathrm{O}_{10+\delta}$ (Bi-2223) exhibits the highest superconducting transition temperature within the Bi-based cuprates\,\cite{Tallon1988,Tatascon1988}, making it especially significant for understanding the mechanisms of high-temperature superconductivity.
Studies of Bi-2223 not only provide insight into the electronic structure and pairing symmetry in multilayer cuprates, but also offer guidance for the design and optimization of high-$T_{\mathrm{c}}$ superconducting materials. Moreover, its potential applications in high-field superconducting magnets, superconducting power transmission, and quantum information technologies underscore the broader relevance of Bi-2223 in the pursuit of novel superconductors.

A defining characteristic of superconductors is the opening of a superconducting energy gap in the electronic density of states \cite{BCS2}. In high-temperature cuprate superconductors, the magnitude of this gap reflects the binding strength of Cooper pairs, while its symmetry provides crucial information about the underlying pairing mechanism \cite{cupratesgap}. 
For example, an isotropic $s$-wave gap is typically associated with phonon-mediated pairing, as described by conventional BCS theory, whereas a $d$-wave gap, characterized by nodes along the Brillouin zone diagonals, suggests that electron-electron interactions dominate the pairing process, rather than a purely phonon-driven mechanism \cite{theoryRVB}. Consequently, determining the symmetry of the superconducting gap is essential for elucidating the microscopic origin of high-temperature superconductivity.

A variety of spectroscopic techniques have been employed to probe the superconducting gap.  
Angle-resolved photoemission spectroscopy (ARPES) provides momentum-resolved access to the electronic band structure and superconducting gap \cite{ARPES2003}. Scanning tunneling microscopy and spectroscopy (STM/STS), by measuring the tunneling current–voltage ($I$–$V$) characteristics of a sample, reveals a distinct energy gap in the superconducting state, reflecting the formation of Cooper pairs and providing information on the gap magnitude and symmetry \cite{Fischer2007}. Quasiparticle interference (QPI) imaging further enables visualization of the Fermi surface and gap anisotropy  \cite{Hoffman2002,McElroy2003,Hanaguri2007}. A prevailing consensus has emerged in favor of $d$-wave symmetry in the cuprates \cite{Cuprate_Rev}, supported by a wide array of measurements including  ARPES\,\cite{ARPES_1993}, STS \cite{STM1995,Hoffman2002,McElroy2003,Hanaguri2007} and phase-sensitive Josephson junctions experiments\,\cite{Josephson_dwave,Wollman1993,Tsuei1994}. However, there also exist Josephson junction experiments \cite{Josephson_1999,Josephson2021} and STS studies \cite{s-wave_STM} that support isotropic $s$-wave pairing. 
These conflicting results have left room for debate, underscoring the need for bulk-sensitive, momentum-resolved probes.

Resonant inelastic X-ray scattering (RIXS) has emerged as a powerful bulk-sensitive technique capable of probing charge, spin, and orbital excitations over a wide region of momentum space \cite{intro_RIXS}. RIXS measures the energy and momentum transferred to the system by incident photons, providing access to low-energy collective excitations. Theoretically, it has been proposed that RIXS could also be sensitive to the phase and amplitude of the superconducting order parameter \cite{theory2013}. However, the small energy scale of the superconducting gap posed a challenge due to previously insufficient energy resolution. Recent advances have improved the energy resolution at the Cu $L$-edge to $\sim$ 30 meV \cite{ESRF2018,Diamond}, making it feasible to directly probe the superconducting gap features. Previous RIXS studies have identified gap-like features in Bi-based cuprates and YBa$_2$Cu$_3$O$_{7-\delta}$ \cite{npj_follow,Giacomo2024}. However, these features persisted above T$_c$, suggesting a relation to the pseudogap rather than superconducting gap. Furthermore, the limited measured momenta in these studies prevented the determination of gap symmetry.

In this work, we use high-resolution Cu $L_3$-edge RIXS to investigate the superconducting gap in the overdoped cuprate Bi$_2$Sr$_2$Ca$_2$Cu$_3$O$_{10+\delta}$ (Bi-2223). By analyzing antisymmetrized spectra and tracking the temperature-dependent low-energy spectral weight across various momentum transfers, we uncover distinct superconducting gap features below T$_c$ and characterize their momentum dependence. To interpret these results, we compute the momentum-dependent charge susceptibility for both $d$-wave and isotropic $s$-wave gap symmetries. Comparison with experimental data reveals that the gap evolution in Bi-2223 is consistent with a $d$-wave symmetry. These findings not only strengthen the case for $d$-wave pairing in trilayer cuprates but also demonstrate the efficacy of RIXS as a momentum-resolved, bulk-sensitive probe of superconducting gap symmetry, offering a promising pathway for studying unconventional superconductivity beyond the limitations of surface-sensitive techniques.

\begin{figure*}[htbp]
\includegraphics[width=\textwidth]{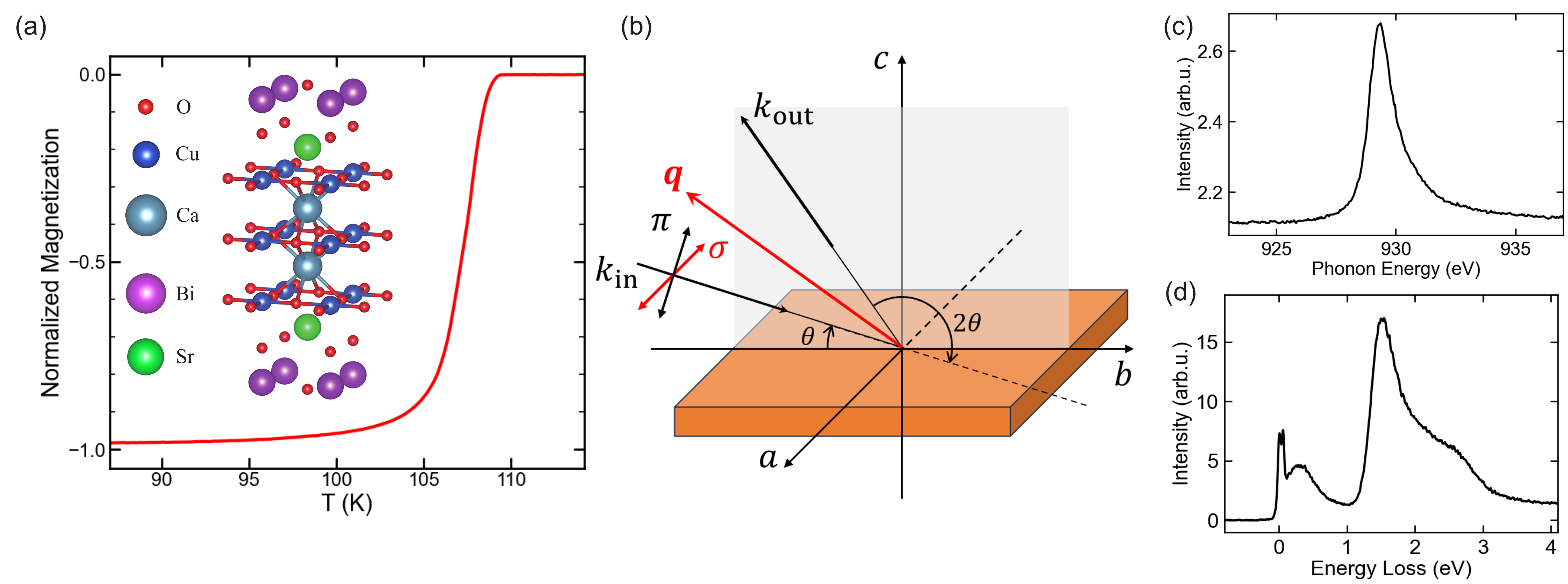}% Here is how to import EPS art
\caption{\label{fig:0} (a) Temperature dependent magnetic susceptibility of Bi-2223, exhibiting a superconducting transition temperature of 107 K. Inset: the crystal structure of Bi-2223 with three Cu-O layers.
(b) Schematic of the experimental geometry. The scattering plane, defined by the incident and outgoing X-ray beams, lies along the crystallographic $a$ or $b$ axis (equivalent in the tetragonal structure). The incident angle is denoted by $\theta$, and $2\theta$ is the angle between the incident and scattered beams. The incident $\pi$-polarization is parallel to the scattering plane, while $\sigma$-polarization is perpendicular to it. The momentum transfer is represented by $\mathbf{q}$; negative $\mathbf{q}$ corresponds to a grazing-incidence configuration in our convention.
(c) Total fluorescence yield (TFY) X-ray absorption spectrum of Bi-2223 at the Cu $L_3$ edge, measured at normal incidence.
(d) Representative RIXS spectrum at $\mathbf{q}_\parallel = (-0.26, 0)$ r.l.u., obtained with a $\sigma$-polarized incident beam.}
\end{figure*}

\section{Material and Method}

We studied an overdoped Bi2223 single crystal with an onset superconducting temperature $T_\mathrm{c} = $107 K and a transition width $\Delta T_\mathrm{c}\approx 3$ K, as shown in Fig. ~\ref{fig:0}(a). The high-quality single crystals were grown by travelling-solvent floating zone method and post-annealed in high-pressure oxygen atmosphere at $550\ ^\circ\mathrm{C}$ for 7 days. Details of the synthesis and post-annealing procedures have been reported previously  \cite{growth2002,BLiang_2004}.  
The selection of an overdoped sample was intentional to suppress the influence of the pseudogap, which is prevalent in underdoped cuprates and reduces the electronic density of states above T$_c$ \cite{Ding1996,Yoshida2009}. This choice allows for a more direct comparison between the normal state and superconducting state in the absence of the pseudogap.

Resonant inelastic X-ray scattering (RIXS) measurements were performed at the Cu $L_3$-edge ($\sim$ 930 eV) using the I21 RIXS beamline at the Diamond Light Source \cite{Diamond}.  The energy resolution was determined by measuring the
full-width at half-maximum of the non-resonant diffuse
scattering from a carbon tape and estimated to be 41.4 meV.  The momentum resolution is estimated to be 0.01 \text{\AA}$^{-1}$ \cite{Diamond}. Figure ~\ref{fig:0}(b) shows the experimental geometry with grazing incidence. Data were collected with the vertical-polarization ($\sigma$-pol) of the incident X-ray, i.e. perpendicular to the scattering plane, to enhance the charge excitations. The scattering $2\theta$ was fixed at 154$^\circ$ and we obtained different momentum transfers within the a-b plane ($\boldsymbol{q}_\parallel$) by changing the $\theta$ angle. We aligned the sample with the Bragg peaks to ensure the direction of the sample prior to the RIXS measurements. Each RIXS spectrum was collected for 12 minutes and normalized by the incident photon flux, and presented in
reciprocal lattice units (r.l.u.) defined  with lattice constants $a=b=3.853 ~\text{\AA}$ and $c=37.072 ~\text{\AA}$ for Bi-2223. 

\begin{figure}[htbp]
\includegraphics[width=0.48\textwidth]{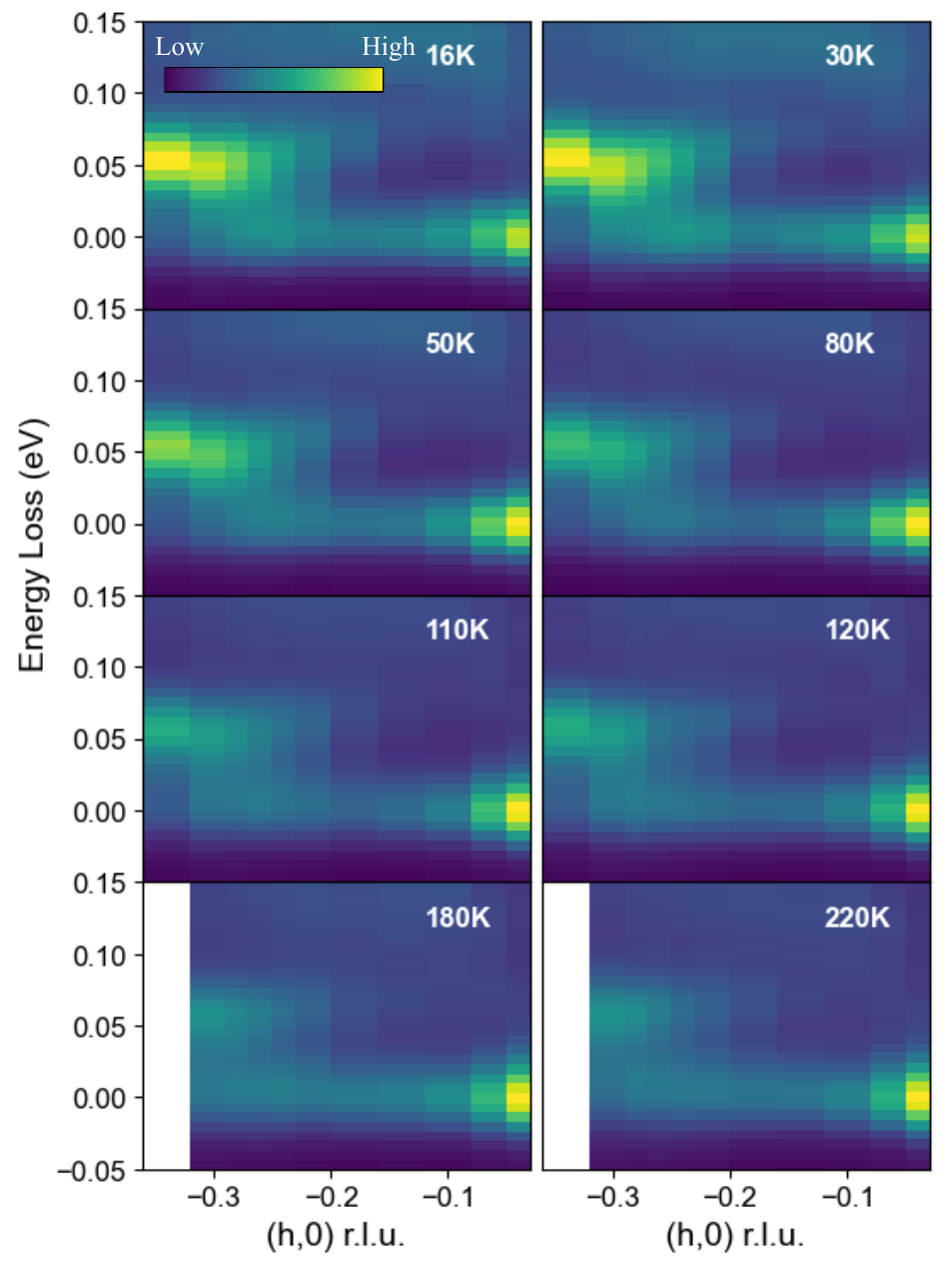}
\caption{\label{fig:1} RIXS intensity map of Bi-2223 as a function
of energy loss and projected in-plane momentum along the Cu-O
bond direction at different temperatures, collected
with a $\sigma$-polarized incident beam. The RIXS spectra were normalized by the incident photon flux and corrected for self-absorption effects. }
\end{figure}

\section{\label{sec:result}Experiment result}

\subsection{RIXS spectrum}

We first measured the Cu $L_3$-edge X-ray absorption spectrum (XAS) using the total fluorescence yield (TFY) method at normal incidence, as shown in Fig.\ref{fig:0}(c). The spectrum exhibits a sharp peak at 929.33 eV, corresponding to the resonant Cu $2p^63d^9 \rightarrow 2p^53d^{10}$ core-level excitation. This energy was selected as the incident energy for the subsequent RIXS measurements to maximize resonant enhancement. A representative RIXS spectrum is shown in Fig.\ref{fig:0}(d), featuring distinct components: the elastic peak at zero energy loss, low-energy phonon and charge excitations below 0.5 eV, and $dd$ orbital excitations in the 1–3 eV range. The zero-energy-loss position was calibrated by fitting the elastic peak to the instrument resolution function. All spectra were corrected for self-absorption effects using established procedures described in Ref.\cite{abcorrection}. The $dd$ excitations remain nearly unchanged with temperature as shown in the Appendex Fig. \ref{dd}, highlighting their localized character. This demonstrates both the sample quality and the stability of the RIXS measurements at different momenta and temperatures.

Figure \ref{fig:1} presents the low-energy momentum-energy intensity maps at different temperatures across T$_c$.
At zero energy loss, we observe a weak but discernible charge density wave (CDW) peak at $\mathbf{q}_\mathrm{CDW} \sim (-0.27, 0)$ r.l.u., consistent with previous observations in overdoped Bi$_2$Sr$_2$CaCu$2$O$_{8+\delta}$ (Bi-2212) \cite{Lu2022}. In addition, well-defined phonon excitations appear at approximately 50 meV, with intensity that increases with momentum transfer. The integrated RIXS intensities in the energy ranges -25 to 25 meV and
25 to 100 meV, corresponding to the quasielastic and phonon-dominated regions, are shown in Appendix Fig.\ref{integrate}, highlighting their distinct spectral distributions. At small momentum transfers, the strong elastic intensity arises from specular reflection. Notably, the elastic background also increases gradually with temperature, in accordance with Bose statistics, as reported in Ref.\cite{Giacomo2024}. The CDW feature already appears at 220 K and becomes more pronounced upon cooling, and the momentum-dependent phonon intensity follows trends consistent with prior studies \cite{Zou2024,Lee2021}.

\begin{figure}[htbp]
\includegraphics[width=0.48\textwidth]{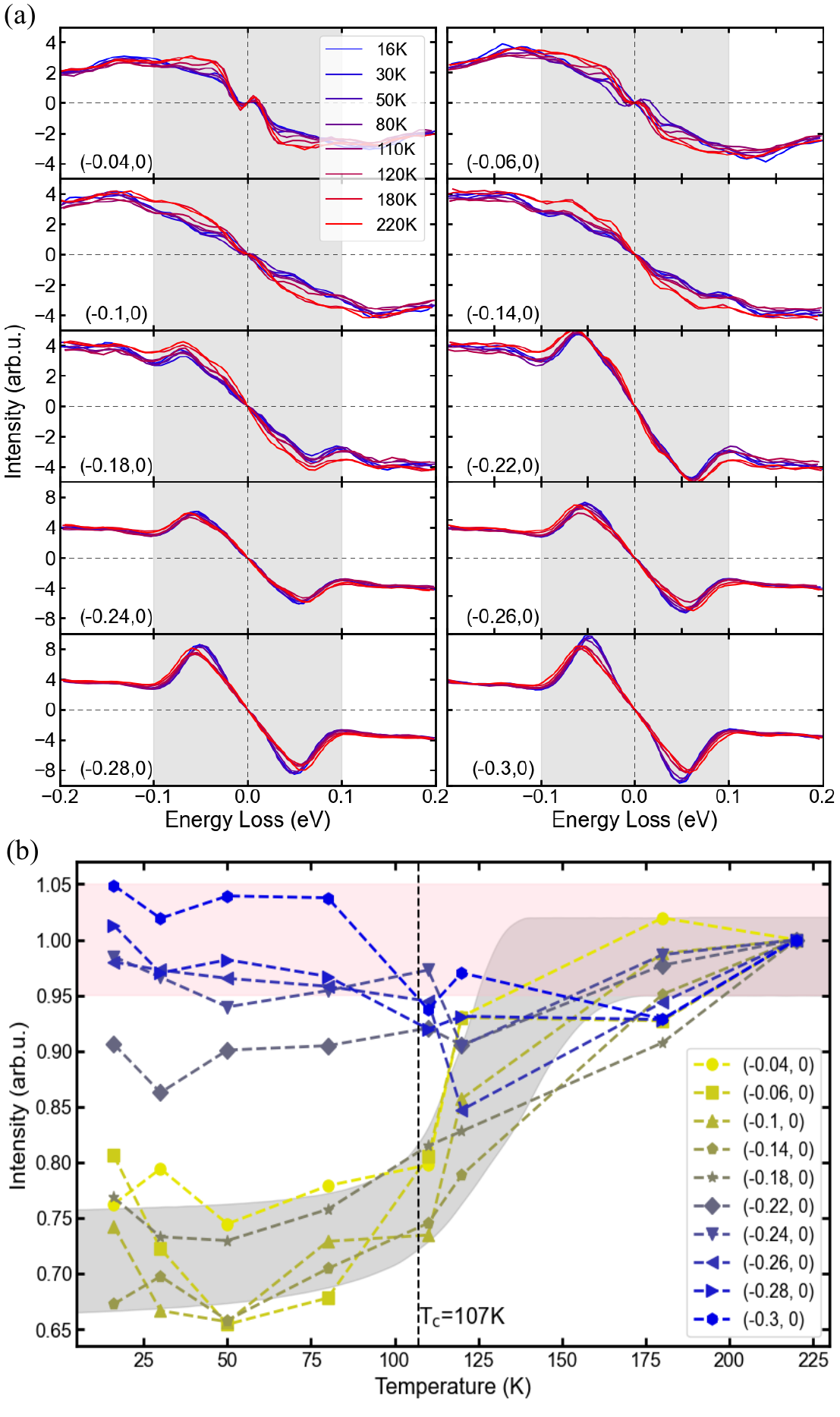}% Here is how to import EPS art
\caption{\label{fig:2}(a) Low-energy antisymmetrized temperature dependent spectra for different momenta indicated in the panels. (b) Integrated intensity of the absolute value of (a) for the grey shaded areas for different momentum transfers, normalized to the 220 K point. The dashed line indicates the T$_c$. }
\end{figure}

\subsection{Observation of superconducting gap}

\begin{figure}[htbp]
\includegraphics[width=0.49\textwidth]{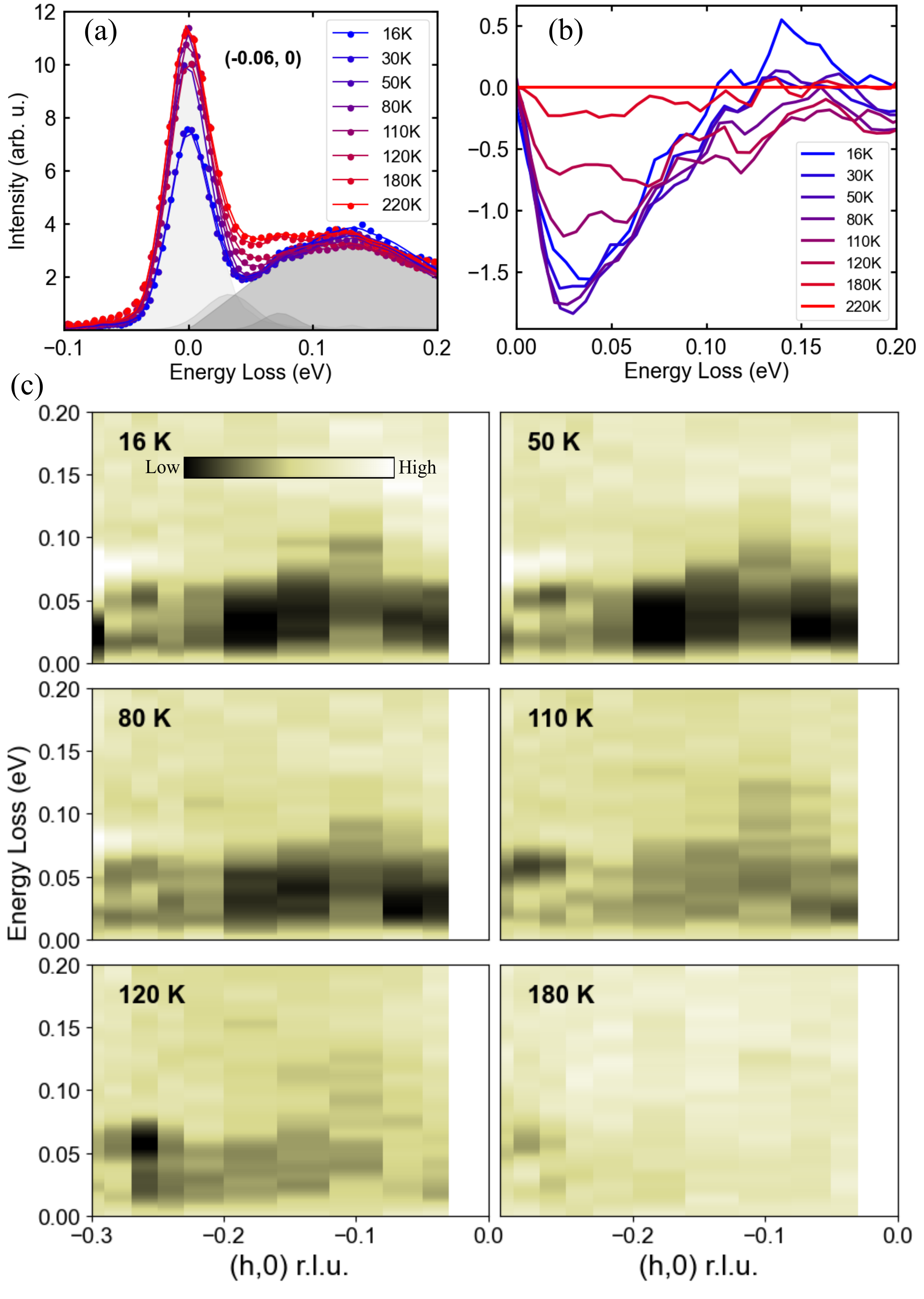}% Here is how to import EPS art
\caption{\label{fig:3} (a) RIXS spectra at $\bf{q}_{\parallel}$ = (-0.06, 0) r.l.u. for Bi-2223. (b) Difference spectra from the highest temperature 220 K. The difference spectra are evaluated after subtracting the phonon and elastic peaks. (c) Intensity map of the difference spectra as a function of
energy loss and projected in-plane momentum along the Cu-O bond direction at different temperatures. }
\end{figure}

\begin{figure*}[htbp]
\includegraphics[width=\textwidth]{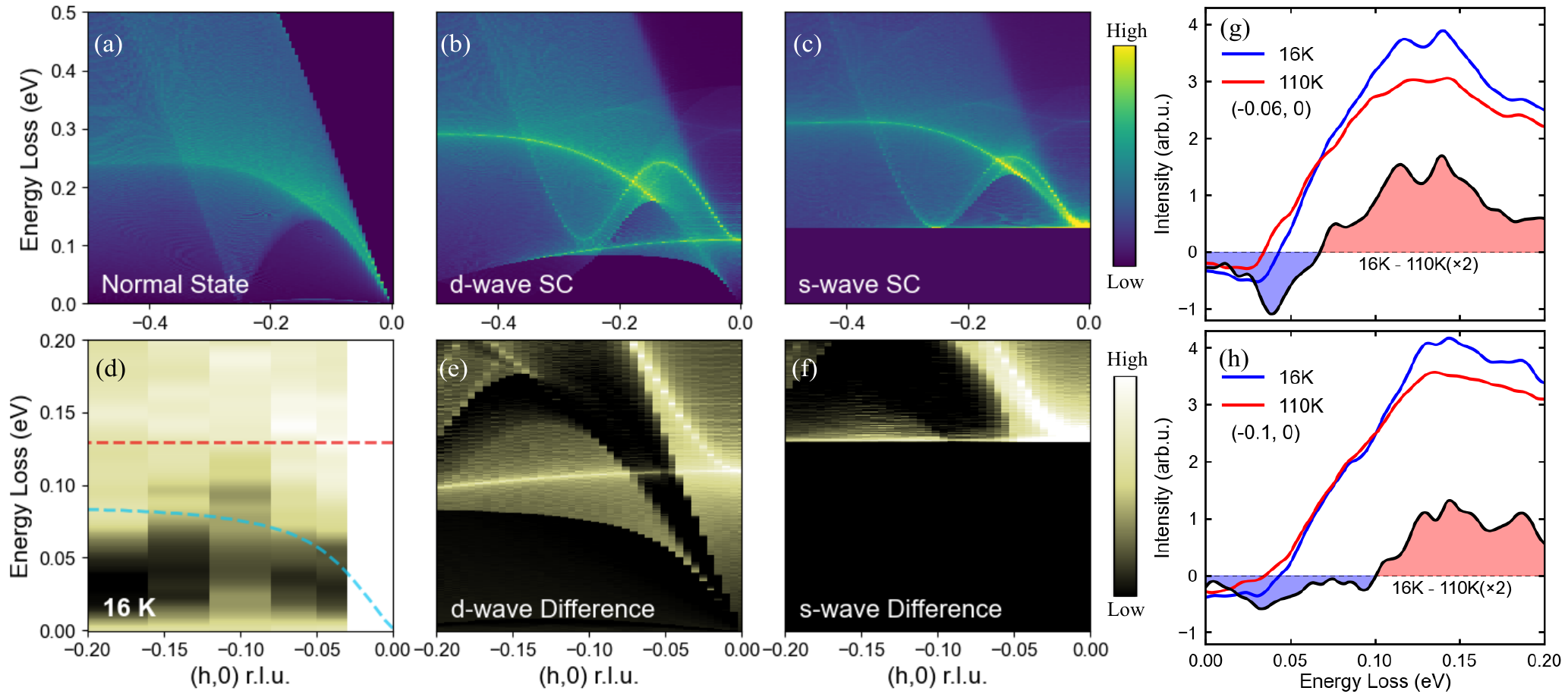}% Here is how to import EPS art
\caption{\label{fig:4}
(a-c) Calculated imaginary part of the charge susceptibility along the Cu–O direction for the normal state (a), $d$-wave superconducting state (b), and $s$-wave superconducting state (c).
(d) Experimental intensity map of the difference spectra at 16 K.
(e, f) Difference in charge susceptibility between the superconducting states of $d$-wave (e) and $s$-wave (f) and the normal state over the energy-loss range 0 – 0.2 eV and momentum transfer from (0,0) to (–0.2, 0) r.l.u.. (g, h) Representative RIXS spectra at $\mathbf{q}_\parallel = (-0.06, 0)$ r.l.u. (g) and (–0.10, 0) r.l.u. (h), after subtraction of the phonon and elastic contributions. Blue and red curves correspond to spectra at 16 K and 110 K, respectively; the black lines show the difference spectra between these temperatures.}
\end{figure*}

To eliminate the influence of the elastic peak, we used the methodology introduced by Merzoni et al.\cite{Giacomo2024} using the integrated results of antisymmetrized data to identify the presence of features related to the gap. We selected the absolute values of the data from -100 meV to 100 meV for integration, obtaining the specific variation of the spectral weight with temperature at different in-plane momentum transfers ($\boldsymbol{q}_\parallel$). For comparison across temperatures, the integrated intensities were normalized to the corresponding value at 220 K, as shown in Fig.~\ref{fig:2}b. The results reveal a clear momentum-dependent suppression of spectral weight with temperature. At small momentum transfers ($|\boldsymbol{q}_\parallel| \leq 0.18$ r.l.u.), the integrated intensity decreases upon cooling below $T_\mathrm{c}$, indicating the formation of a superconducting gap. In contrast, no appreciable spectral change is observed at larger momentum transfers ($|\boldsymbol{q}_\parallel| \geq 0.22$ r.l.u.) due to the pronounced phonon contribution.

Figure~\ref{fig:3}(a) presents representative temperature-dependent RIXS spectra at $\bf{q}_{\parallel}$ = (-0.06, 0) r.l.u.. A clear suppression of the spectral weight is observed below approximately 0.1 eV as the temperature decreases. To analyze this evolution in more detail, we fit the spectra using four components: an elastic peak, a buckling phonon near 35 meV, a breathing phonon around 60 meV, and a broad high-energy feature composed of bimagnon excitations and low-energy interband transitions. The elastic peak was fitted using the instrumental resolution function, phonons were fitted with resolution-limited Gaussian profiles, and high-energy excitation was captured using an anti-Lorentzian function, consistent with previous studies \cite{YY2018,YY2019}.
To extract the temperature-induced spectral changes, the intensity, width, and position of the buckling phonon were fixed to their values at 220 K. The breathing phonon's intensity and width were also kept constant, with its energy allowed to vary with temperature. Additional fitting details are provided in Appendix Fig.~\ref{fitting}. 
We emphasize that this subtraction and fitting approach is reliable at low momentum transfers, as previous RIXS studies have shown that phonon energies and intensities remain temperature independent  when it is away from the CDW ordering vector
$\bf{q}_{\mathrm{CDW}}$ = (-0.27, 0) r.l.u. \cite{Lee2021,Lu2022,Zou2024}. Near
$\bf{q}_{\mathrm{CDW}}$, phonon softening and intensity enhancement complicate the analysis due to the appearance of the CDW order. 
Figure~\ref{fig:3}(b) displays the difference spectra obtained by subtracting the 200 K spectrum from those at lower temperatures. These reveal a systematic suppression of the low-energy spectral weight below $T_\mathrm{c}$, accompanied by the formation of a dip within the gap region, which is indicative of the opening of the gap. The two-dimensional energy–momentum map in Fig.~\ref{fig:3}(c) further illustrates this behavior. Gap-related suppression is confined to small $\boldsymbol{q}_\parallel$, while additional low-energy spectral features appear at larger momenta due to the influence of the CDW order. The gap becomes increasingly pronounced at lower temperatures, which is consistent with the development of the superconducting order.

\subsection{Charge susceptibility and gap symmetry}

To gain further insight into the gap symmetry, we calculate the charge susceptibility $\chi_c(\bf{q},\omega)$ using a single band model in both superconducting and normal states, where $\bf{q}$ and $\omega$ denote momentum transfer and energy loss from the X-ray photons to the sample, respectively. The RIXS intensity is proportional to the imaginary part of the charge susceptibility, which can be expressed as \cite{theory2013}
{\small \begin{eqnarray}
&&\mathrm{Im}\,\chi(\mathbf{q},\omega)\propto \label{suscep}\\\nonumber
&&\sum_{k}\left[1+\frac{\mathrm{Re}(\Delta_{\mathbf{k}}\Delta^*_{\mathbf{k+q}})-\varepsilon_\mathbf{k}\varepsilon_\mathbf{k+q}}{E_\mathbf{k}E_\mathbf{k+q}}\right]\times\delta(\omega-E_\mathbf{k}-E_\mathbf{k+q})
\end{eqnarray}}
where $E_\mathbf{k}=\sqrt{\varepsilon_\mathbf{k}^2+\Delta_\mathbf{k}^2}$ is the quasiparticle energy dispersion, and 
$\varepsilon_\mathbf{k}$ is the bare band dispersion using a tight-binding approximation and $\Delta_\mathbf{k}$ is the superconducting (SC) gap function. The hopping parameters are taken from ARPES measurements in Ref. \cite{luo2023electronic}, and the dispersion is given by $\varepsilon_\mathbf{k}=-2t[\cos(k_x a)+\cos(k_y a)]-4t'\cos(k_x a)\cos(k_y a)-2t''[\cos(2k_x a)+\cos(2k_y a)]-\mu$, with parameters $t=0.123$, $t'=-0.37t$, $t''=-0.47t'$ and $\mu=-0.092$. In the normal state, the superconducting gap is set to zero ($\Delta_k\equiv 0$). 
In the superconducting state, we consider (i) an isotropic s-wave pairing and (ii) a d-wave symmetry pairing: $\Delta_\mathbf{k}\equiv\Delta_0$
for the s-wave, and $\Delta_\mathbf{k}=\Delta_0[\cos(k_x a)-\cos(k_y a)]/2$ for the d-wave, with gap amplitude $\Delta_0=65\ \mathrm{meV}$ \cite{luo2023electronic}. The tri-layer Bi2223 has inequivalent inner and outer CuO$_2$ planes, here we use the superconducting gap value of the inner planes, which exhibit a larger gap than the outer ones \cite{luo2023electronic}. Figure \ref{fig:4} shows the imaginary parts of the charge susceptibility along the (h,0) direction. 
In the superconducting state, creating the particle-hole excitations requires an energy larger than the SC gap size, leading to a suppression of the low-energy spectral weight relative to the normal state. For the s-wave gap, the spectral weight vanishes entirely within the range 0 $<$ $\omega$ $<$ 2$\Delta_0$, reflecting the isotropic nature of the gap. In contrast, the d-wave state shows a narrower region of vanishing spectral weight, with residual intensity at the $\Gamma$ point due to the vanishing of the SC gap along the nodal direction. This constitutes a key distinction between the two gap symmetries.

At high energy-loss regions, the spectral weight is similar between the d-wave and s-wave gap symmetries. However, significant differences appear below 2$\Delta_0$ ($\sim$ 130 meV). To facilitate a direct comparison with the experiment, we plot the difference in charge susceptibility between the d-wave and s-wave cases, after subtracting the normal-state response [Fig. \ref{fig:4}(e, f)]. This can be directly compared with the experimental spectral weight measured at 16 K after subtracting the 220 K data [Fig.\ref{fig:4} (d)]. For visual clarity, we overlay the contours of the theoretical s-wave and d-wave gaps on the experimental data.
In the energy range where the s-wave charge susceptibility is fully suppressed, the d-wave case retains substantial spectral intensity, consistent with the experimental observations. In particular, the presence of finite spectral weight at the
$\Gamma$ point in the d-wave case agrees well with the measured data. Moreover, a peak-and-dip feature appears in the
difference RIXS spectra between 110K and 16K [Fig.\ref{fig:4}(g, h)], indicating the spectral weight transferred from lower energy to higher energy loss below T$_c$. The peak resembles the 2$\Delta_0$ peak observed in the Raman data of Bi2223 \cite{npj_follow}. The gap size increases with momentum, in agreement with the theoretical prediction for a d-wave superconducting gap. 
These results strongly support the d-wave symmetry of the superconducting order parameter in Bi-2223 and demonstrate the capability of RIXS to probe the gap symmetry in high-temperature superconductors.

\section{Discussion}

Recent high-resolution laser-based ARPES studies on overdoped Bi-2223 with a superconducting transition temperature T$_c$ of 108 K have resolved the splitting of three distinct bands arising from the three CuO$_2$ layers: the $\alpha$ and $\beta$ bands associated with the outer CuO$_2$ layers (OP), and the $\gamma$ band associated with the inner CuO$_2$ layer (IP) \cite{luo2023electronic}. The momentum transfer 
\textbf{q}= (-0.18, 0) r.l.u., below which the superconducting gap becomes visible in the RIXS spectra, approximately connects the antinodal regions of the $\beta$-band Fermi surface ($\Delta k_F \sim$ 0.2 r.l.u.), rather than those of the $\alpha$ or $\gamma$ bands. This observation may be attributed to the stronger spectral weight of the $\beta$ band near the antinode compared to the $\alpha$ and $\gamma$ bands, as reported in ARPES measurements \cite{luo2023electronic}.  ARPES studies on Bi-2223 have mapped the superconducting gap along the Fermi surface and confirmed its momentum dependence to be consistent with d-wave symmetry \cite{ARPES2002, ARPES1, luo2023electronic, ARPES2025}. Similarly, scanning tunneling microscopy (STM) studies have observed a characteristic V-shaped in the $\mathrm{d}I/\mathrm{d}V$ curve, further supporting a nodal d-wave gap structure in Bi-2223 \cite{STM2009, STM2022, STM2020}. Our momentum- and temperature-dependent RIXS results are consistent with these ARPES and STM findings, reinforcing the identification of d-wave symmetry in the superconducting gap of Bi-2223.

Despite these promising results, there are notable limitations in using RIXS to probe superconducting gap symmetry. First, the current energy resolution limits the applicability of the technique to systems with relatively large superconducting gaps. However, ongoing instrumental developments are addressing this constraint. For example, the new RIXS beamline BL02U at NanoTerasu has achieved energy resolutions as high as $\sim$ 9.4 meV at the oxygen $K$-edge and $\sim$ 16 meV at the Cu $L_3$-edge \cite{Terasu2}, allowing gap studies in a wider class of superconductors. 
Second, our results indicate that at the present level of energy resolution, RIXS predominantly detects the largest superconducting gap in materials with multigap structures. In Bi-2223, ARPES has revealed three superconducting gaps associated with the $\alpha$, $\beta$, and $\gamma$ Fermi surfaces, with maximum gap values of 17 meV, 29 meV, and 62 meV, respectively \cite{luo2023electronic}. Our main analysis employs a single band model focusing on the largest gap associated with the inner CuO$_2$ layer. 
To assess the impact of the multigap nature, we also calculated the charge susceptibility using a three-band model, which includes the largest gap from inner-plane pairing ($\Delta_{\mathrm{ip}}^0=65\ \mathrm{meV}$), a moderate gap from outer-plane pairing ($\Delta_{\mathrm{op}}^0=20\ \mathrm{meV}$), and a smaller interlayer pairing gap between the outer planes ($\Delta_{\mathrm{oo}}^0=5\ \mathrm{meV}$). Additional model details and parameters are provided in Ref.~\cite{luo2023electronic}. 
As shown in Appendix Fig.~\ref{triT}, the three-band model yields a charge susceptibility with significantly smaller gap features compared to the one-band model. This discrepancy arises from the smaller pairing gaps in the outer layers and the effects of interlayer coupling. 
Remarkably, the experimental RIXS spectra align more closely with the single-band model predictions, indicating that RIXS is most sensitive to the largest superconducting gap in multiband systems. This preferential sensitivity underscores both the strength and the current limitation of the technique: While it enables direct access to the largest pairing energy, it may obscure smaller-scale features associated with secondary bands. 
We attribute this limitation to the present data statistics and instrumental resolution. With future enhancements in photon flux and energy resolution, it should become possible to resolve finer spectral structures, as anticipated from theoretical calculations of the charge susceptibility.

\section{Conclusion}

In summary, we employed high-resolution Cu L$_3$-edge RIXS to investigate the superconducting gap in overdoped Bi-2223 across a range of momentum transfers and temperatures. Our measurements reveal a superconducting gap (2$\Delta_0 \sim$ 130 meV) that emerges below $T_\mathrm{c}$ and exhibits a pronounced momentum dependence, with spectral weight suppression occurring primarily at small momentum transfers ($|\boldsymbol{q}_\parallel| \leq 0.18$ r.l.u.).
By comparing the experimental RIXS spectra with theoretical calculations of the momentum-resolved charge susceptibility for both d-wave and isotropic s-wave pairing symmetries, we find that the observed gap behavior is consistent with a d-wave superconducting order parameter. These results reinforce prior ARPES and STM findings while providing bulk-sensitive, momentum-resolved evidence for d-wave pairing in the trilayer cuprate Bi-2223. Our work demonstrates the power of RIXS as a complementary probe to surface-sensitive techniques, enabling direct access to superconducting gap features in systems with complex or inaccessible surfaces. As RIXS instrumentation continues to advance in energy resolution, this approach holds significant potential for probing the superconducting gap structure in a broad class of unconventional superconductors.

\vspace{1 ex}
\begin{acknowledgments}
\noindent
We acknowledge the valuable discussions with Thomas P. Devearaux, Giacomo Ghiringhelli and Atsushi Fujimori. The RIXS experimental data were collected at beamline I21 of the Diamond Light Source in Harwell Campus, United Kingdom. Y.Y.P. is grateful for financial support from the Ministry of Science and Technology of China (Grants No. 2021YFA1401903 and No. 2024YFA1408702), the National Natural Science Foundation of China (Grant No. 12374143), and Beijing Natural Science Foundation (Grant No. JQ24001). X.J.Z’s work is supported by the National Natural Science Foundation of China (Grant No. 12488201) and the National Key Research and Development Program of China (Grant No. 2021YFA1401800). Y.L. acknowledges the financial support from the Ministry of Science and Technology of China (Grant No. 2022YFA1403000) and the National Natural Science Foundation of China (Grant No. 12274207).
\end{acknowledgments}

\section{Appendix}

\subsection{Momentum and temperature dependence of dd excitations}

\begin{figure}[htbp]
  \centering
  \includegraphics[width=0.45\textwidth]{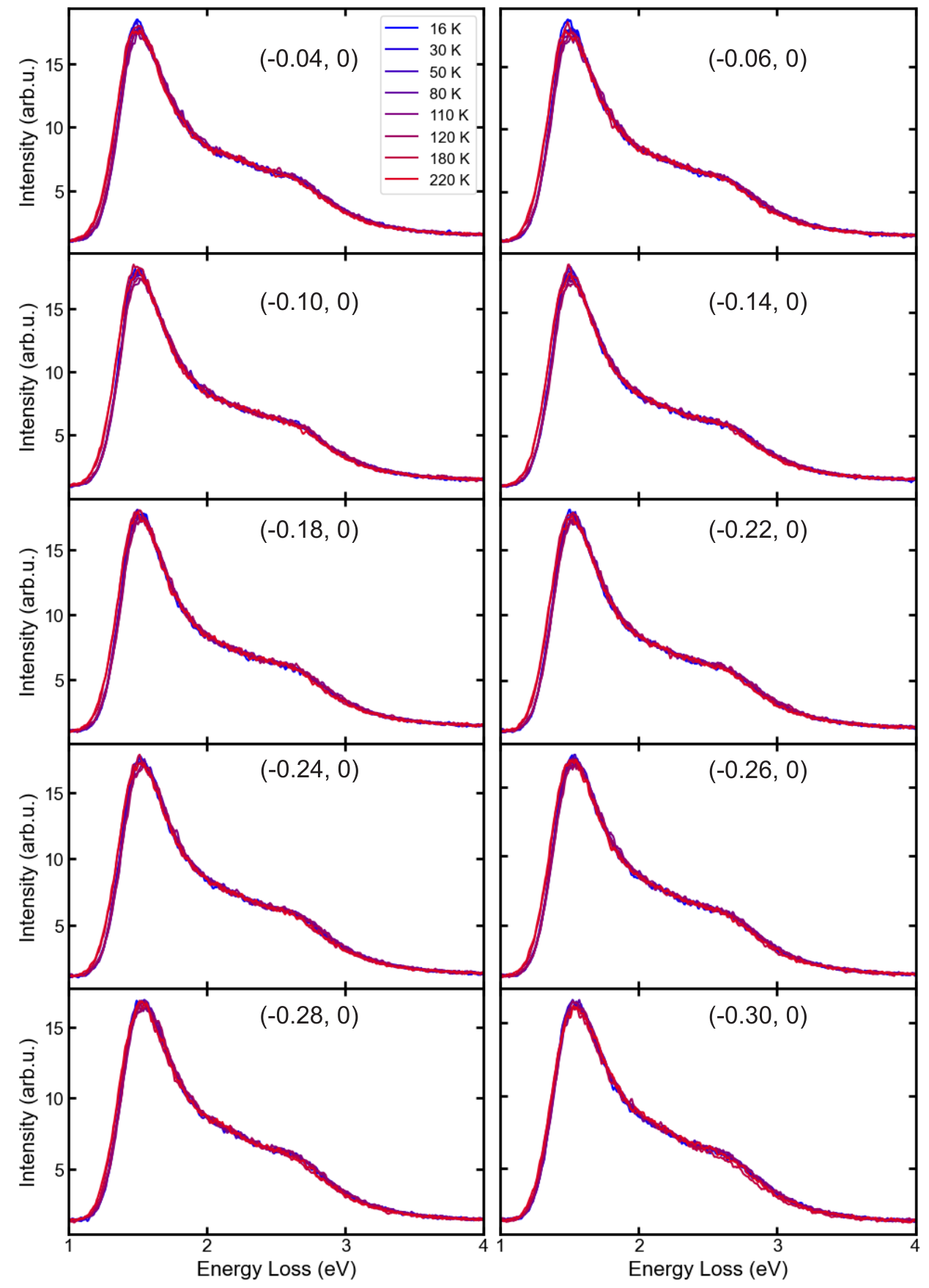}
  \caption{Temperature dependence of RIXS spectra showing the $dd$ excitations at different temperatures and momentum-transfers.}
  \label{dd}
\end{figure}

\begin{figure}[htbp]
  \centering
  \includegraphics[width=0.42\textwidth]{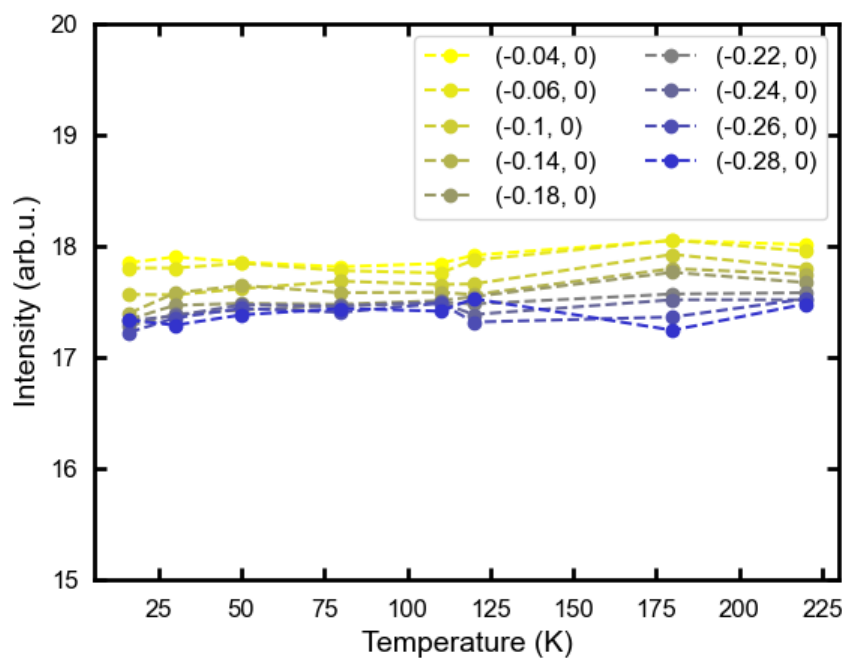}
  \caption{Temperature dependence of integrated intensity from 1 to 4 eV energy loss for $dd$ excitations.}
  \label{dd_inte}
\end{figure}

The temperature dependence of the high-energy $dd$ excitations at various momentum transfers is presented in Fig. \ref{dd}. The spectral line shapes exhibit no discernible change as a function of temperature. This indicates that the $dd$ excitations are insensitive to the superconducting transition. Quantitatively, the integrated spectral weight in the 1–4 eV energy window, shown in Fig. \ref{dd_inte}, remains essentially constant with temperature at all momenta investigated. This robustness of the
$dd$ excitations against temperature variation highlights the selectivity of RIXS in probing different electronic degrees of freedom, with high-energy excitations serving as an internal reference when analyzing temperature-dependent spectral weight redistribution at low energies.

\subsection{Low energy features and the fit of the RIXS spectra}

\begin{figure}[htbp]
  \centering
  \includegraphics[width=0.4\textwidth]{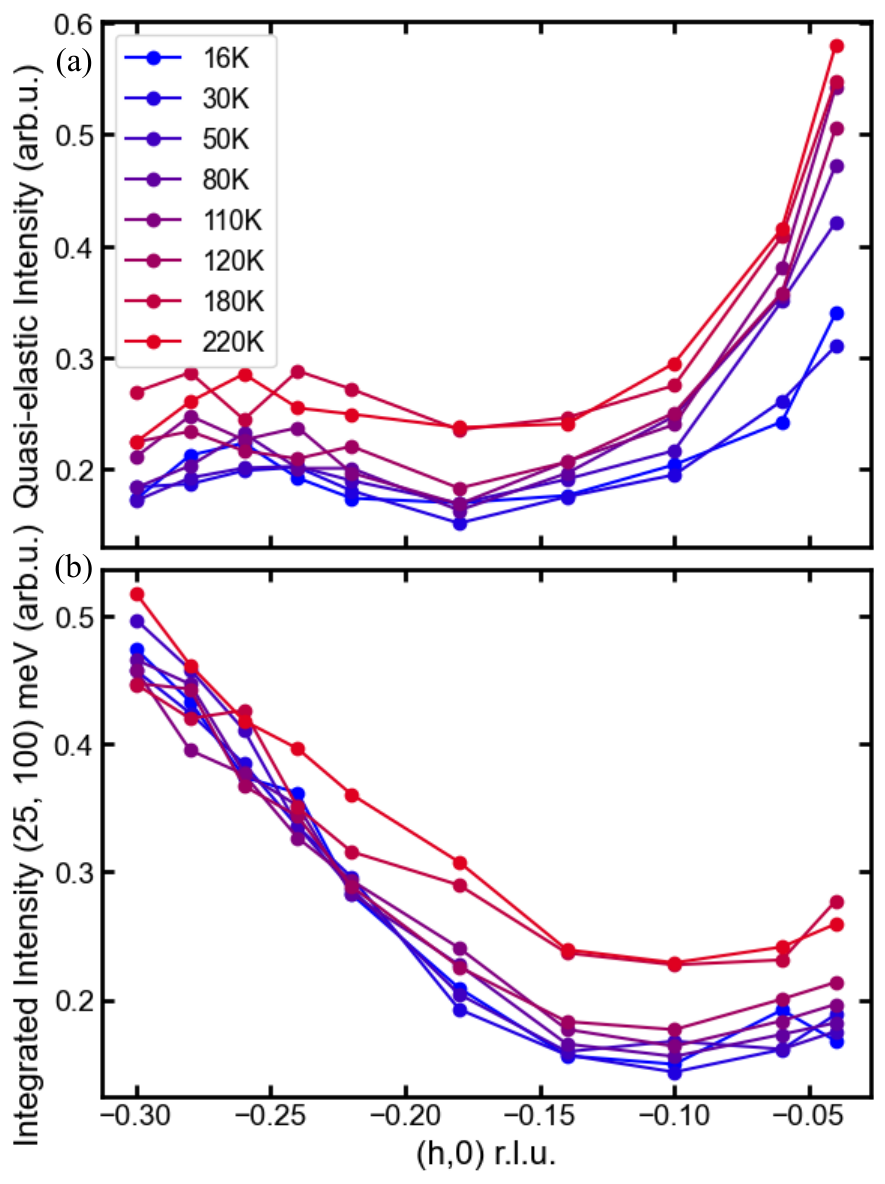}
  \caption{(a) Intensity at 0 $\pm$ 25 meV for the quasielastic signal. (b) Integrated intensity from 25 to 100 meV for phonon-dominated signals. Both panels illustrate the momentum dependence of the intensity at different temperatures.}
  \label{integrate}
\end{figure}

Figure~\ref{integrate} shows the integrated RIXS intensities in the energy ranges -25 to 25 meV and
25 to 100 meV, corresponding to the quasielastic and phonon-dominated regions, respectively. The momentum dependence of these integrated intensities was tracked across the superconducting transition temperature
$T_{\mathrm{c}}$. At small momentum transfers, the strong quasi-elastic intensity originates from the tail of the specular angle by surface reflectivity effects. Superimposed on this background, we observe a weak, broad CDW feature near
(-0.27, 0) r.l.u. that becomes more pronounced as the temperature is lowered. In contrast, the integrated intensity in the 25 to 100 meV range, dominated by phonon excitations, exhibits a systematic increase with momentum transfer, reflecting the stronger contribution of phonons at larger \textbf{q}. Importantly, the data reveal a clear suppression of low-energy spectral weight below $T_{\mathrm{c}}$, indicating the formation of a superconducting gap that depletes low-energy electronic excitations. This suppression is most significant at small momentum transfers, where the superconducting gap exerts the strongest influence on the low-energy spectral response.

\begin{figure}[htbp]
  \centering  \includegraphics[width=0.45\textwidth]{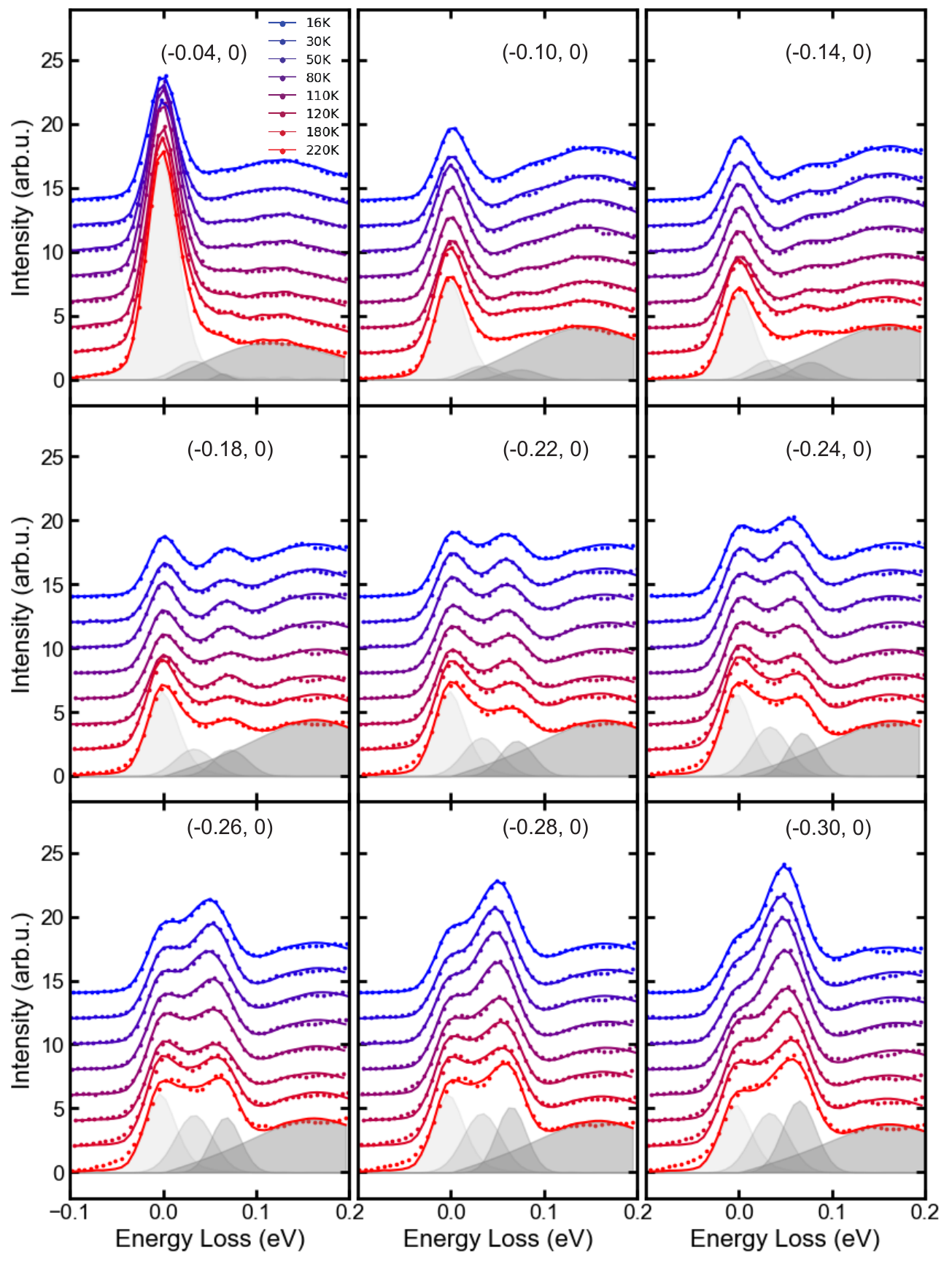}
  \caption{Fitting of the low-energy RIXS spectra at various temperatures and momentum transfers. The fit includes the instrumental resolution function for the elastic peak, two resolution-limited Gaussian functions for the buckling phonon ($\sim$ 33 meV) and the breathing phonon (50–80 meV), and an anti-Lorentzian function for the high-energy background. The total fit is shown as solid lines. Spectra are vertically offset for clarity.}
  \label{fitting}
\end{figure}

Figure \ref{fitting} presents the RIXS spectra measured at various temperatures and momentum transfers. As the momentum transfer increases, the phonon signal becomes more pronounced, while the elastic peak diminishes in intensity. We model the low-energy spectral region using four components: the instrumental resolution function for the elastic peak, two resolution-limited Gaussian peaks corresponding to the breathing mode ($\sim$ 33 meV) and buckling phonon mode (50-80 meV), and an anti-Lorentzian function describing the higher-energy excitations, following the fitting procedure in ref. \cite{Zou2024}. An example fit to the data at 220 K is shown, demonstrating excellent agreement between the fitted curves and the experimental spectra.

\subsection{Calculation of charge susceptibility with three-band models}

\begin{figure*}[htbp]
  \centering
  \includegraphics[width=\textwidth]{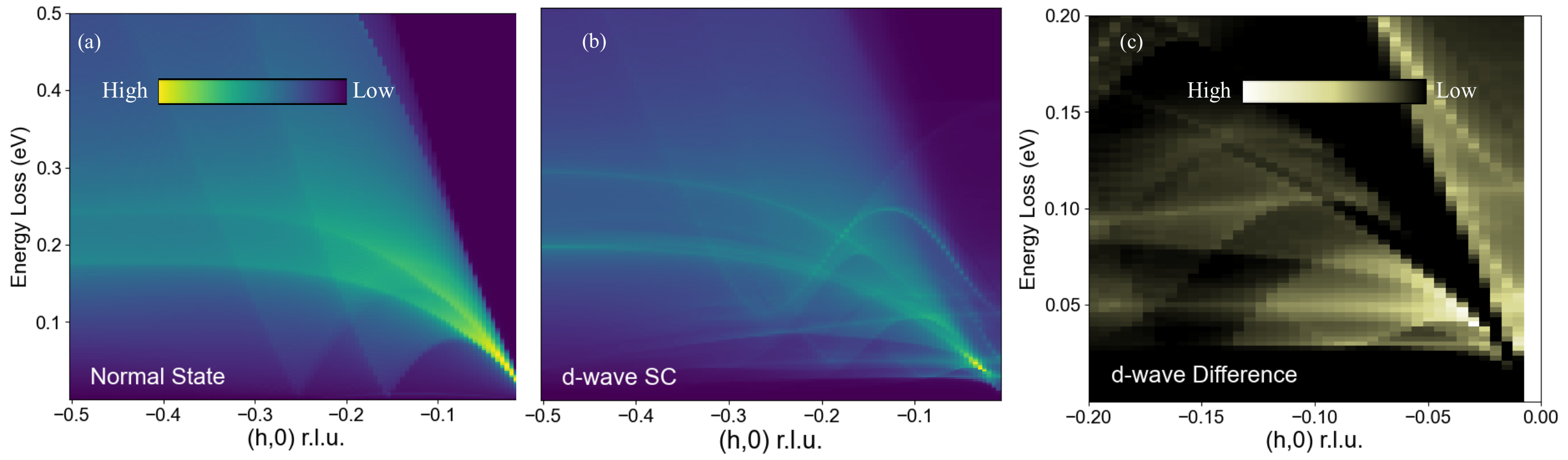}
  \caption{Calculated imaginary part of the charge susceptibility with three-band model along the Cu–O direction for the normal state (a) and $d$-wave superconducting state (b). (c) Difference in charge susceptibility between the superconducting states of $d$-wave and the normal state over the energy-loss range 0 – 0.2 eV and momentum transfer from (0,0) to (–0.2, 0) r.l.u.. }
  \label{triT}
\end{figure*}

The charge susceptibility using a three-band model is expressed as 
{\small \begin{eqnarray}
&&\mathrm{Im}\,\chi(\mathbf{q},\omega)\propto\\\nonumber
&&\sum_{k}\int_{-\infty}^{0} d\omega^{\prime}\mathrm{Tr}[\Gamma A(\mathbf{k},\omega^{\prime})\Gamma A(\mathbf{k+q},\omega^{\prime}+\omega)]\\\nonumber
&&\times[f(\omega^{\prime})-f(\omega^{\prime}+\omega)]
\end{eqnarray}}
where $f(\omega)$ is the Fermi-Dirac distribution. Under the zero-temperature approximation, $f(\omega)$ reduces to the Heaviside function $\theta(-\omega)$. The wave vector $\textbf{k}$ spans the full Brillouin zone and $\omega$ denotes the energy loss from the X-ray photons to the sample, whereas $\omega'$ corresponds to an intermediate energy. The vertex function $\Gamma$ corresponds to the density vertex operator, given as $\mathrm{diag}(1,1,1,-1,-1,-1)$ in the Nambu basis $\Psi^{\dagger}_{\mathbf{k}} = (c^{\dagger}_{\mathbf{k},1},c^{\dagger}_{\mathbf{k},2},c^{\dagger}_{\mathbf{k},3},c_{-\mathbf{k},1},c_{-\mathbf{k},2},c_{-\mathbf{k},3})$. The spectral function is defined by
$A(k,\omega)=-\dfrac{1}{\pi}\mathrm{Im}G(\mathbf{k},\omega+i\eta)$, where the Green's function $G(\mathbf{k},\omega+i\eta)=[\omega+i\eta-H(\mathbf{k})]^{-1}$. The Hamiltonian $H(\mathbf{k})$ is represented as a 6 $\times$ 6 matrix incorporating both intra- and interlayer couplings as well as pairing terms, which can be explicitly written as
\begin{equation}
    H(\mathbf{k})=\begin{pmatrix}
    \epsilon_{op} & t_{io} & t_{oo} & \Delta_{op} & 0 & \Delta_{oo}\\
    t_{io} & \epsilon_{ip} & t_{io} & 0 & \Delta_{ip} & 0\\
    t_{oo} & t_{io} & \epsilon_{op} & \Delta_{oo} & 0 & \Delta_{op}\\
    \Delta_{op} & 0 & \Delta_{oo} & -\epsilon_{op} & -t_{io} & -t_{oo}\\
    0 & \Delta_{ip} & 0 & -t_{io} & -\epsilon_{ip} & -t_{io}\\
    \Delta_{oo} & 0 & \Delta_{op} & -t_{oo} & -t_{io} & -\epsilon_{op}
    \end{pmatrix}
\end{equation}

The model parameters were adopted from the ARPES measurements in Ref. \cite{luo2023electronic}. The calculated charge susceptibility is presented in Fig.\ref{triT}, providing a direct comparison between the three-band model and the single-band results shown in Fig. 5(a,b). In the normal state [Fig.\ref{triT}(a)], the inclusion of the outer CuO$_2$ layer band introduces additional spectral features arising from interband transitions, absent in the single-band calculation. In the superconducting state [Fig.\ref{triT}(b)], the spectral weight in the gap region is modified by the smaller pairing gap in the outer layers and by interlayer coupling effects, which broaden and redistribute the spectral weight. As a result, the difference in charge susceptibility after subtracting the normal-state response exhibited a narrower energy range in the suppression of spectral weight [Fig.\ref{triT}(c)], inconsistent with the larger gap region observed in the experimental data [Fig.~5(d)].

%\vspace{5cm}
\vspace*{8cm}

\providecommand{\noopsort}[1]{}\providecommand{\singleletter}[1]{#1}%
%

%\bibliography{reference}% Produces the bibliography via BibTeX.

\end{document}